# Transferable Machine Learning Approach for Predicting Electronic Structures of Charged Defects

Yuxing Ma[†,1,2], Yang Zhong[†,1,2], Yu Hongyu[1,2], Shiyou Chen[1,2], Hongjun Xiang[*,1,2]

[1]Key Laboratory of Computational Physical Sciences (Ministry of Education), Institute of Computational Physical Sciences, State Key Laboratory of Surface Physics, and Department of Physics, Fudan University, Shanghai, 200433, China

[2]Shanghai Qi Zhi Institute, Shanghai, 200030, China

[†]These authors contribute equally to this work.

*E-mail: hxiang@fudan.edu.cn

**Abstract:** The study of the electronic properties of charged defects is crucial for our understanding of various electrical properties of materials. However, the high computational cost of density functional theory (DFT) hinders the research on large defect models. In this study, we present an E(3) equivariant graph neural network framework (HamGNN-Q), which can predict the tight-binding Hamiltonian matrices for various defect types with different charges using only one set of network parameters. By incorporating features of background charge into the element representation, HamGNN-Q enables a direct mapping from structure and background charge to the electronic Hamiltonian matrix of charged defect systems without DFT calculation. We demonstrate the model's high precision and transferability through testing on GaAs systems with various charged defect configurations. Our approach provides a practical solution for accelerating charged defect electronic structure calculations and advancing the design of materials with tailored electronic properties.

## I. Introduction

Recently, the integration of machine learning (ML) techniques and density functional theory (DFT) has fostered advancements in computational materials science. For instance, ML has made significant progress in prediction for the physical and chemical properties of materials[1–5], inverse design[6–10], and accelerating molecular dynamics simulation[11–15]. Generally, DFT requires many costly self-consistent iterations to get converged Hamiltonian matrices, resulting in huge computational costs for systems with a large number of atoms. This limitation hinders the study of material properties in large systems at the DFT precision level. ML methods have powerful fitting capabilities and can accelerate this process.

Currently, obtaining the electronic structure of materials still relies on time-consuming DFT calculations, which have poor scalability. Especially when dealing with defective systems, due to the periodic boundary conditions and the interaction between defects, a large unit cell needs to be considered, which increases the computational cost. To solve this problem, semi-empirical tight-binding (TB) approximations[16] such as the Slater-Koster method[17] are proposed. These methods can reduce the computation requirements compared to the DFT method, but they cannot accurately predict the electronic structure of general systems. Therefore, the acceleration of the electronic structure calculation process using ML methods is desired. Although some studies have considered the equivariance of the Hamiltonian in their studies[18,19], graph neural network models based on deep learning exhibit superior accuracy and generalization capabilities[20,21], especially when driven by big data sets. Notably, graph neural networks based on E(3) equivariance are regarded as the current optimal prediction method and have successfully been applied to predict the tight-binding Hamiltonian of both periodic and non-periodic systems, including bulk materials, two-dimensional materials, and organic molecules [22–24].

The presence of defects in materials is inevitable in the synthesis processes of materials. Defects play a crucial role in determining the properties and performance of materials, including their electronic structure, electrical, and optical properties[25–28]. Charged defects are particularly important as well since they introduce non-equilibrium charge states into materials, thereby altering their electronic structure and properties[29–32]. Therefore, studying the behavior and properties of charged defects in semiconductor materials is of significant importance for understanding the performance of semiconductor devices and optimizing device designs. However, there is currently a lack of an ML model that can be used to accelerate the DFT calculation of charged defects. Therefore, the precise and rapid determination of the electronic structure of charged defect materials is of significance in advancing the material design.

In this study, we designed a HamGNN-Q model for predicting the ab initio TB Hamiltonian of charged defect system. The features of background charge (Q) and element representation (Z) are integrated to realize the transfer of charge information by the message-passing neural network (MPNN) mechanism. By fitting the TB Hamiltonian of charged defects obtained from first-principles calculations, we accelerated the computation of the electronic structure for systems with charged defects. Through training on GaAs systems with charged defects, we achieved high-precision predictions of the Hamiltonian for different cell sizes and various defect configurations with the same network weights, indicating a high level of transferability and generalization ability. Our model provides a practical approach for accelerating electronic structure calculations in large-scale systems with charged defects.

## II. Method

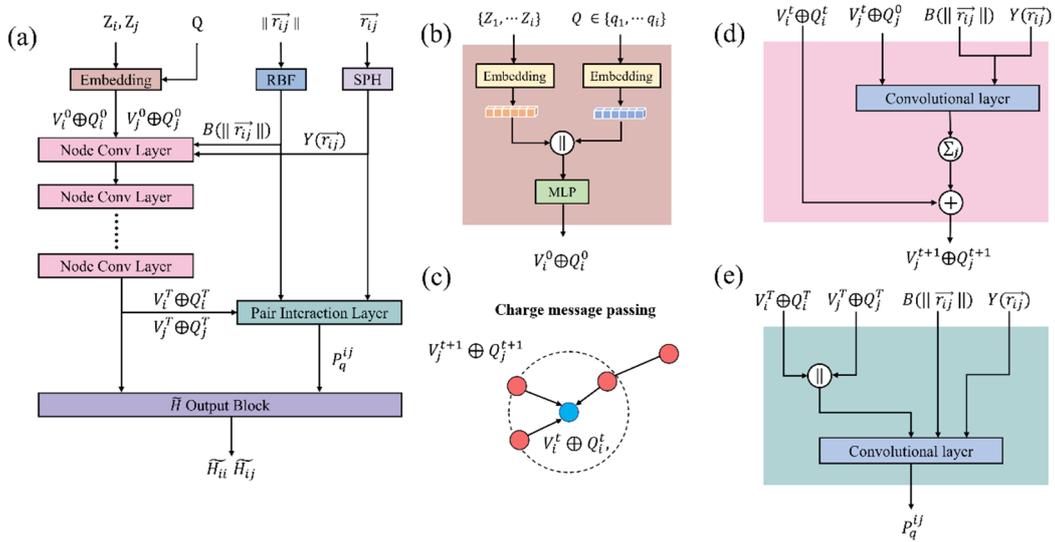

**Fig.1** (a) HamGNN-Q architecture. (b) Schematic diagram of background charge embedding. (c) The illustration of the charge message passing. (d) Node conv layer. (e) Pair interaction layer.

The network architecture of HamGNN-Q is shown in Fig. 1(a). HamGNN-Q can realize direct mapping from atomic species $Z_i$ and atomic positions $\vec{R}_i$ to ab initio

TB Hamiltonian matrix. HamGNN-Q first embeds the atom types $Z_i$, interatomic distances $\|\vec{r}_{ij}\|$, and relative orientations $\vec{r}_{ij}$ in the initial feature vectors $V_i^0$. To embed charge information into the model and thus fit the Hamiltonian of charged defects, we introduce the background charge feature Q. The atomic number and the background charge are initialized as one-hot vectors $V_i^0$ and $Q_i^0$ respectively. and merged them into the feature vector $V_i^0 \oplus Q_i^0$ by a multi-layer perceptron (MLP), as shown in Fig. 1(b).

The interatomic distances are expanded using Bessel basis functions:

$$B(\|\vec{r}_{ij}\|) = \sqrt{\frac{2}{r_c}} \frac{\sin(n\pi\|\vec{r}_{ij}\|)}{\|\vec{r}_{ij}\|} f_c(\|\vec{r}_{ij}\|), \tag{1}$$

where $f_c$ is the cosine cutoff function. The direction information between atom $i$ and atom $j$ is embedded into a set of real spherical harmonics $Y(\vec{r}_{ij})$, which are used to construct rotational equivariant filters in equivariant message passing. Each node convolutional layer is constructed by using the spherical harmonic basis of $r_{ij}$ and the tensor product of $V_i^T \oplus Q_i^T$, see Eq (2):

$$m_i^{t+1} = \sum_j M_t(V_i^t \oplus Q_i^t, V_j^t \oplus Q_j^t, \vec{r}_{ij}), \tag{2}$$

where $V_i^t$ and $Q_i^t$ denotes the feature of the element and background charges of atom $i$ in the node convolution layer $t$ ($t \leq T$), respectively. Each atom combines its features with the features of its neighboring atoms to generate new messages function $M_t$. The output message $m_i^{t+1}$ is comprised of contributions with corresponding parity to adhere to parity symmetry.

As illustrated in Figure 1(c), HamGNN-Q utilizes MPNN to update node information, in which each atom adjusts its state based on the received message and its current state:

$$V_i^{t+1} \oplus Q_i^{t+1} = U_t(V_i^t \oplus Q_i^t, m_i^{t+1}), \tag{3}$$

where $U_t$ is the update function. Furthermore, in each subsequent iteration, the network progressively integrates and assimilates information from neighboring nodes.

Residual layers adjust the interaction characteristics based on the orbital characteristics of the two interacting atoms and the strength of their orientation. Finally, HamGNN-Q utilizes the equivariant irreducible spherical tensors (EIST) to construct the on-site and off-site Hamiltonian matrices by separating and incorporating node features and interaction features through the $\widetilde{H}$ output block.

For the training of HamGNN-Q, we propose a training method termed as "reciprocal space enhanced training procedure". To increase transferability and avoid overfitting, we include the error of the energy bands as a regularization term in the loss function:

$$L = \left\| \widetilde{H} - H \right\| + \frac{\lambda}{N_{orb} \times N_k} \sum_{k=1}^{N_k} \sum_{n=1}^{N_{orb}} \left\| \widetilde{\varepsilon_{nk}} - \varepsilon_{nk} \right\|, \tag{4}$$

where $L$ represents the total loss function, $\widetilde{H}$ represents the corresponding prediction, $H$ represents the target Hamiltonian calculated from scratch, $\lambda$ represents the loss weight with band energy error, and $N_k$ denotes the number of randomly generated K points in the reciprocal space used for diagonalizing the Hamiltonian in each training step.

The total loss function can be divided into two components: the first term represents the Hamiltonian loss in real space, while the second term represents the band energy loss in reciprocal space. If the predicted Hamiltonian has not yet reached a relatively low error, there may be a significant discrepancy in the predicted energy bands. Therefore, the training of the model is divided into two steps. First, only the mean absolute error (MAE) of the Hamiltonian matrix is used as the loss value to train the network. When the network converges to a certain accuracy, the MAE of each band energy calculated at random K points in the reciprocal space is added to the loss function, and the training is restarted until the network weights are completely converged.

## III. Result

The GaAs system with charged defects was chosen to verify the performance of HamGNN-Q. GaAs demonstrates exceptional optoelectronic properties and semiconductor behavior, making it highly versatile for applications in optoelectronic devices, including solar cells, detector, and LEDs[33–35]. However, during the fabrication process of GaAs, numerous charged defects, such as vacancies, impurities, and dislocations, can potentially arise, exerting significant influences on the performance of GaAs semiconductor materials[36–38]. Therefore, researching the electronic properties of charged defect systems in GaAs is paramount.

The structures with different elemental compositions, various types of defects and amounts of background charge are considered in our data set, as illustrated in Table 1. Through OpenMX calculations, the Hamiltonian matrices corresponding to the structures were obtained to construct the dataset.

As shown in Fig.2 (a), the predicted Hamiltonian matrix exhibits excellent agreement with the calculation by DFT. The trained HamGNN-Q model achieves an MAE of around meV for the Hamiltonian matrix elements on the test set. Furthermore, as depicted in Figure 2(b), the predicted Hamiltonian successfully reproduces the energy levels obtained from DFT calculations, indicating the high accuracy of the model.

**Table 1** The type of defect and the amount of charge in the data set

| Charge | Perfect | $V_{As}$ | $V_{Ga}$ | $I_{As}$ | $I_{Ga}$ |
|---|---|---|---|---|---|
| 0 | √ | √ | √ | √ | √ |
| -3 |   | √ |   |   | √ |
| 3 |   |   | √ | √ |   |

*The black check mark indicates the presence of the respective defect in the dataset.

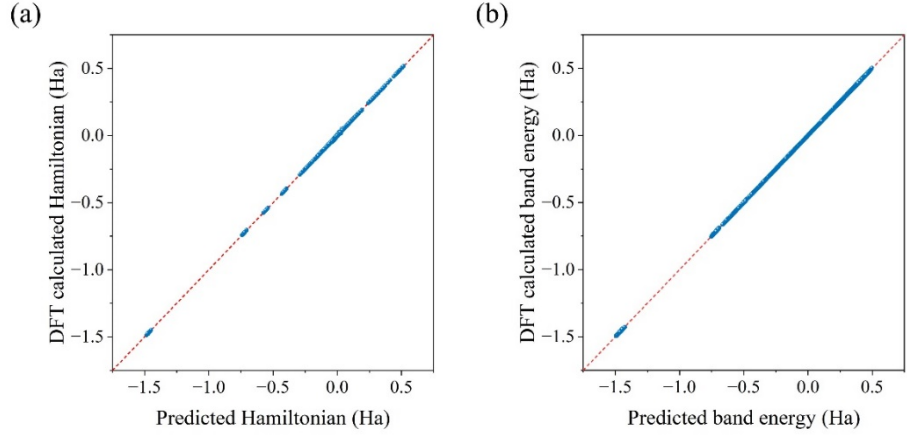

**Fig.2** (a) Comparison of the HamGNN-Q predicted Hamiltonian matrix elements with the OpenMX calculated Hamiltonian matrix elements on the test set. (b) Comparison of predicted and calculated band energy on the test set.

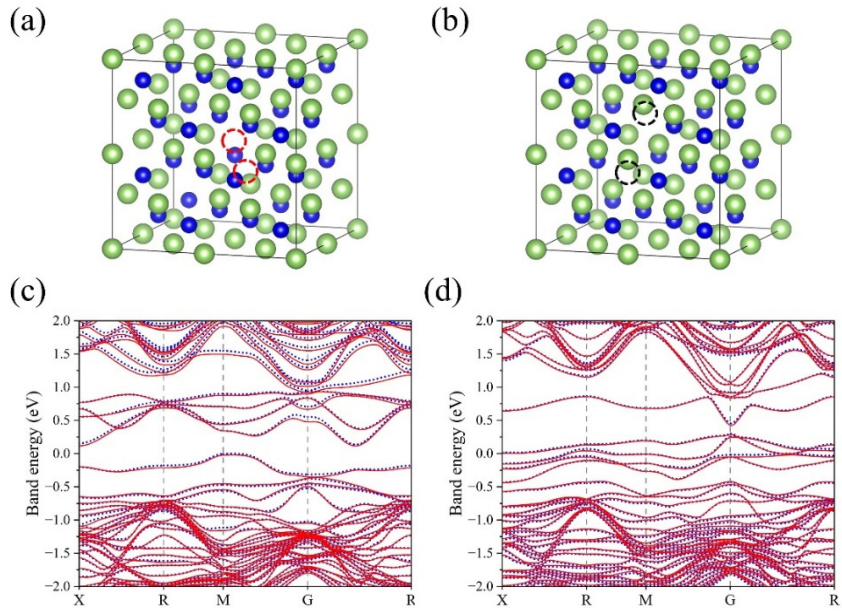

**Fig.3** (a) Crystal structure of nearest-neighbor Ga double-vacancy defect ($DV_{Ga-Ga}^{6-}$). The red dotted circle indicates Ga vacancy. (b) Crystal structures of nearest-neighbor As double-vacancy defect ($DV_{As-As}^{6+}$). The black dotted circle indicates As vacancy. (c-d) The band structure for $DV_{Ga-Ga}^{6-}$ and $DV_{As-As}^{6+}$ computed by HamGNN-Q (red solid line) and DFT calculation (blue short dot), respectively.

In practice, multiple types of defects often coexist in materials. To test the transferability of HamGNN-Q, we constructed different combinations of vacancy and interstitial defects for testing. These structures were not included in the training dataset. As shown in Fig.3 (a-b), we constructed the nearest-neighbor Ga double-vacancy defect ($DV_{Ga-Ga}^{6-}$) and As double-vacancy defect ($DV_{As-As}^{6+}$), respectively. Similarly, Fig.4 (a-b) demonstrates the atomic structure of Ga double interstitial defects ($DI_{Ga-Ga}^{6+}$) and As double interstitial defects ($DI_{As-As}^{6-}$). In both cases, by using the predicted Hamiltonian, we can obtain the band structures that almost exactly agree with the DFT calculations, as depicted in Fig. 3 (c-d) and Fig. 4 (c-d), respectively.

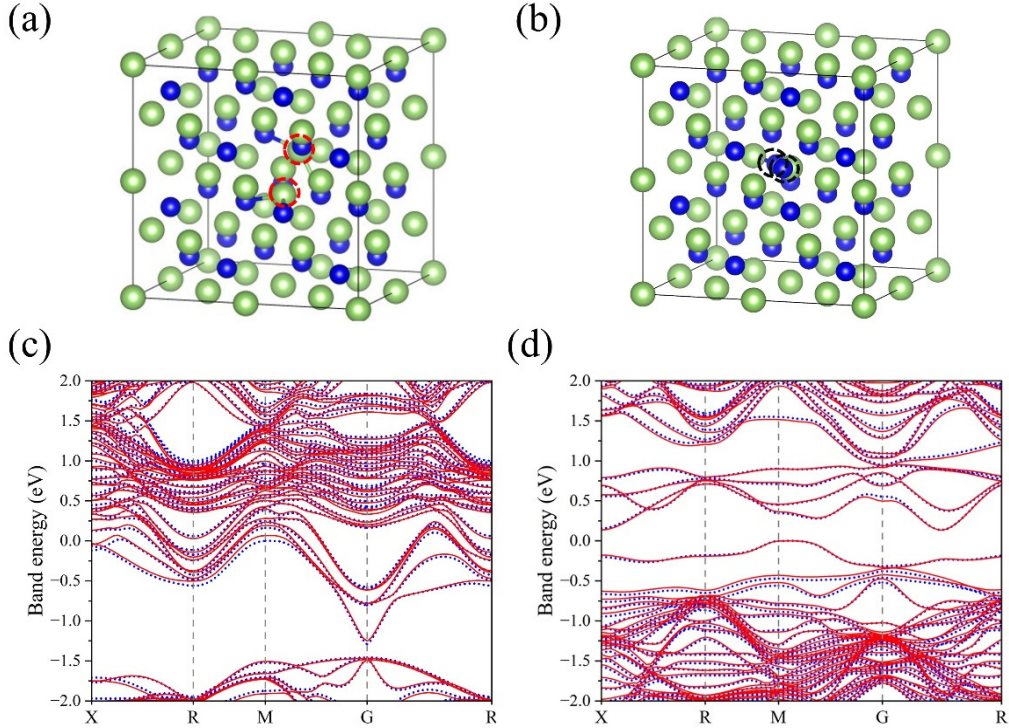

**Fig.4** (a) Crystal structure of Ga double interstitial defects ($DI_{Ga-Ga}^{6+}$). The red dotted circle indicates Ga interstitial defect. (b) Crystal structure of As double interstitial defects ($DI_{As-As}^{6-}$). The black dotted circle indicates As interstitial defect. (c-d) The band structure for $DI_{Ga-Ga}^{6+}$ and $DI_{As-As}^{6-}$ computed by HamGNN-Q (red solid line) and DFT calculation (blue short dot), respectively.

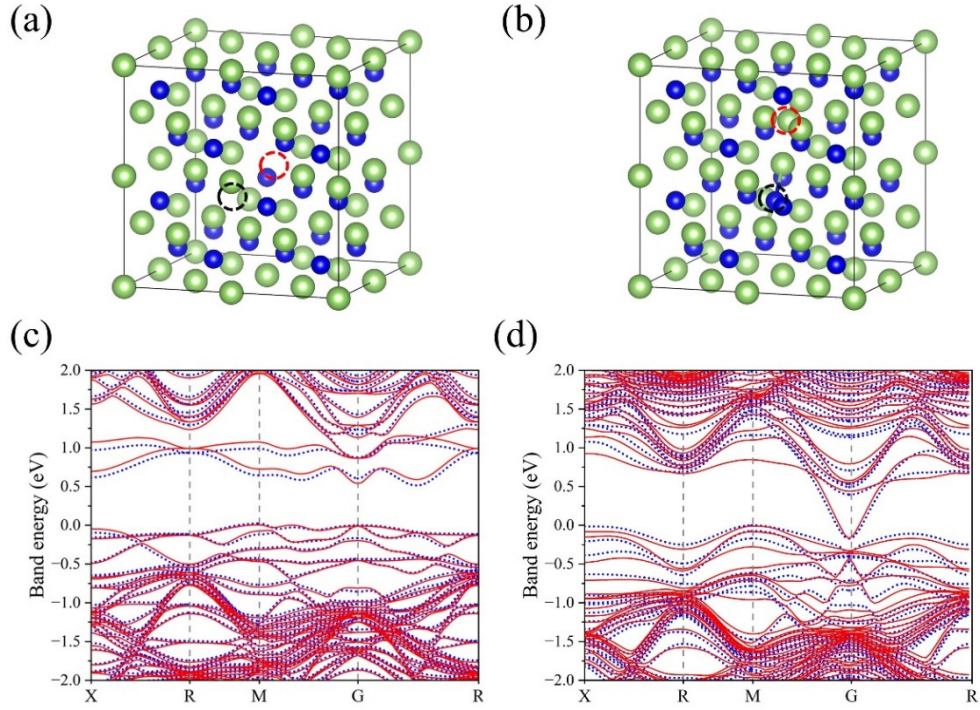

**Fig.5** (a) Crystal structures of the combination of a Ga single-vacancy defect and its nearest neighboring As single-vacancy defect ($DV^0_{Ga-As}$). The red dotted circle indicates Ga vacancy and the black dotted circle indicates As vacancy. (b) Crystal structures of the combination of a Ga interstitial defect and an As interstitial defect ($DI^0_{Ga-As}$). The red dotted circle indicates Ga interstitial defect and the black dotted circle indicates As interstitial defect. (c-d) The band structure for $DV^0_{Ga-As}$ and $DI^0_{Ga-As}$ computed by HamGNN-Q (red solid line) and DFT calculation (blue short dot), respectively.

Moreover, we also investigated the coexistence of different defect types. As shown in Fig.5 (a-b), we constructed the combination of a Ga single-vacancy defect and its nearest neighboring As single-vacancy defect($DV^0_{Ga-As}$) and the combination of Ga interstitial defect and As interstitial defect($DI^0_{Ga-As}$). By employing the predicted Hamiltonian, we calculated the corresponding band structures, as depicted in Fig. 5 (c-d). The obtained results exhibit a good agreement with the DFT calculations, further validating the accuracy and extensibility of our approach.

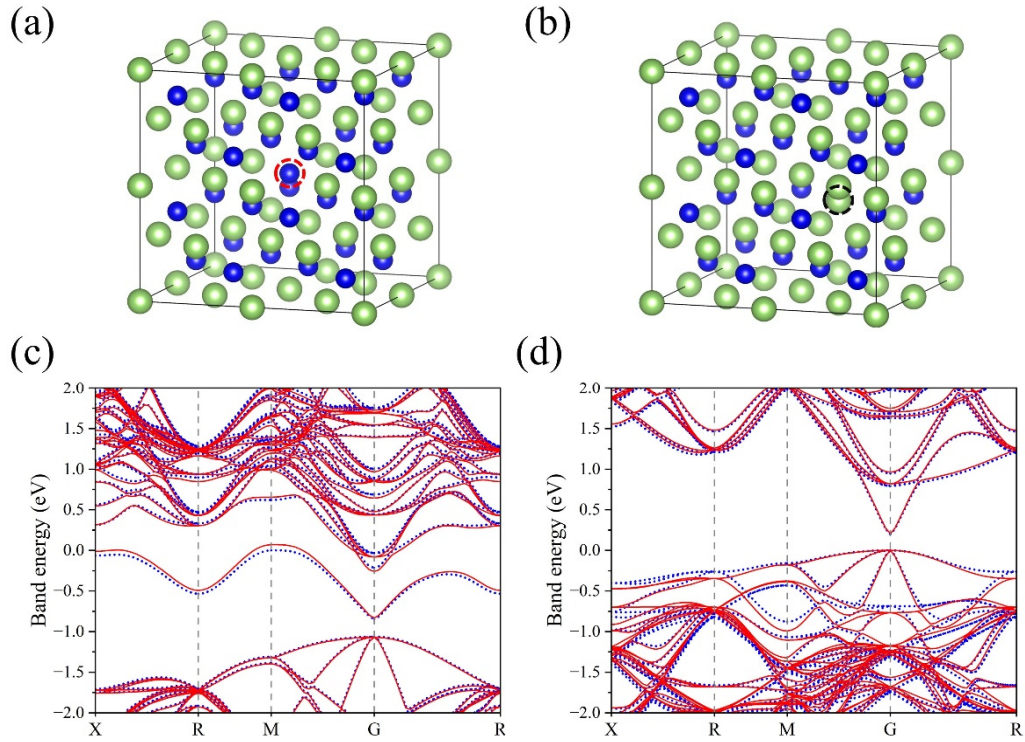

**Fig.6** (a) Crystal structure of Ga substituted by As ($As_{Ga}^{3-}$). The red dotted circle indicates the As atom after replacing Ga. (b) Crystal structure of As substituted by Ga ($Ga_{As}^{3+}$). The black dotted circle indicates the Ga atom after replacing As. (c-d) The band structure for $As_{Ga}^{3-}$ and $Ga_{As}^{3+}$ computed by HamGNN-Q (red solid line) and DFT calculation (blue short dot), respectively.

In addition to the aforementioned cases, we also investigated scenarios involving atomic substitutions. As shown in Fig.6 (a-b), we constructed structures of a single Ga substituted by As ($As_{Ga}^{3-}$) and a single As substituted by Ga ($Ga_{As}^{3+}$), respectively. The band structures obtained from the predicted Hamiltonian exhibit a high agreement with the DFT calculations.

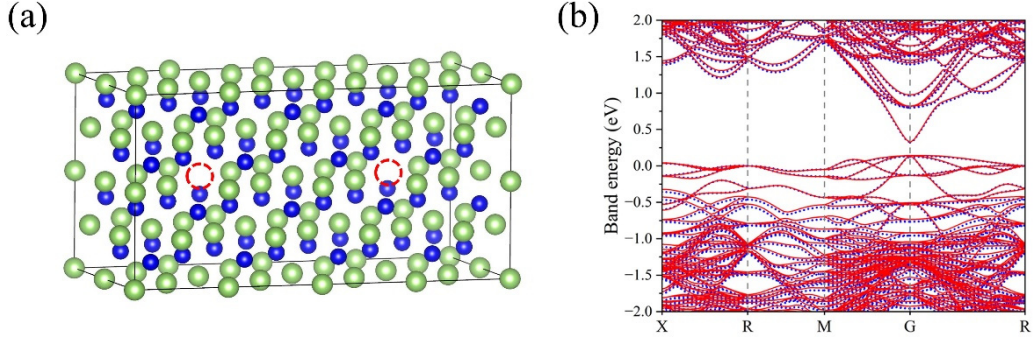

**Fig.7** (a) Crystal structures of Ga double vacancies in 4×2×2 supercell [$DV_{Ga-Ga}^{6-}(sc)$]. The red dotted circle indicates Ga vacancy. (b) The band structure for $DV_{Ga-Ga}^{6-}(sc)$ computed by HamGNN-Q (red solid line) and DFT calculation (blue short dot), respectively.

To verify the prediction performance of the HamGNN-Q for large-sized cells, we constructed a 4×2×2 supercell for testing. It contains two Ga single-vacancy defects that are separated by a certain distance ($DV_{Ga-Ga}^{6-}(sc)$), the crystal structure is shown in Fig.7(a). By utilizing the predicted Hamiltonian from HamGNN-Q, we obtained band structures that are nearly identical to those obtained from DFT calculations, as shown in Fig.7(b). We have thus demonstrated the high efficiency of HamGNN-Q in handling large-scale material systems.

## IV. Summary

In this work, we propose an E(3) equivariant graph neural network for predicting the ab initio TB Hamiltonian (HamGNN-Q), which enables the Hamiltonian of the charged defect system to be fitted by embedding the background charge information into the graph representation. For GaAs charged defect system, our model facilitates a direct correspondence from the structure and charge to the self-consistent Hamiltonian without expensive DFT cost and enables electronic structure computations for large-scale systems with charged defect configurations. Furthermore, the HamGNN-Q demonstrates the remarkable capability of prediction for structures that do not exist in the data set and the interaction between the combinations of different types of defects, which also reflects excellent transferability. HamGNN-Q also demonstrates excellent integration capability by using a single model to fit the Hamiltonian of various point defects with different charges.